\documentclass[12pt]{article}
\usepackage{amsmath,amssymb,amsthm,amsxtra,overpic,bbm,bm,epsfig,subfigure}
\usepackage{hyperref}
\usepackage{color}
\usepackage{stackengine} 
\usepackage{multirow}
\usepackage{slashed}
\usepackage{booktabs} 

\begin{document}
\vspace{0.2cm}

\begin{center}
{\Large\bf Neutrino reheating predictions with non-thermal leptogenesis}\\
\vspace{0.2cm}
\end{center}
\vspace{0.2cm}

\begin{center}
{\bf Xinyi Zhang}~$^{a,b,c}$~\footnote{Email: zhangxy@hbu.edu.cn}
\\
\vspace{0.2cm}
$^a$ {\em \small Department of Physics, Hebei University, Baoding, 071002, China}\\
$^b$ {\em \small Hebei Key Laboratory of High-precision Computation and Application of Quantum Field Theory, Baoding, 071002, China}\\
$^c$ {\em \small Hebei Research Center of the Basic Discipline for Computational Physics, Baoding, 071002, China}
\end{center}

\vspace{1.5cm} 
 
\begin{abstract}
Connecting inflation with neutrino physics through non-thermal leptogenesis via direct inflaton-right-handed neutrino (RHN) coupling naturally incorporates neutrino reheating, leaving no ambiguity regarding the early history of the universe. In ref.~\cite{Zhang:2023oyo}, we demonstrate that non-thermal leptogenesis from inflaton decay expands the viable parameter space compared to thermal leptogenesis and provides a natural link to inflation. In this work, we refine our previous findings by closely examining the dynamics of neutrino reheating. We first calculate the duration of neutrino reheating on a general basis, then analyze inflationary observables consistent with neutrino reheating across four models, establishing a direct connection between baryon asymmetry and the spectral index. This approach places these two important observables on the same plane and yields specific predictions that help break the degeneracy among inflationary models. The well-motivated and economical framework offers a simple, natural, and testable description of the early universe.
\end{abstract}

\section{Introduction}
There is growing evidence that the early universe undergoes an exponentially accelerated expansion era, known as inflation, which addresses both the flatness and horizon problems\cite{Guth:1980zm}. Inflation also seeds the anisotropy observed in the cosmic microwave background (CMB) and baryon acoustic oscillations through quantum fluctuations. For reviews, see, e.g., refs. \cite{Bassett:2005xm, Baumann:2009ds}. In its simplest realization, the single-field slow-roll models, this accelerated expansion is driven by the potential energy of the inflaton, which rolls slowly towards the potential minimum.
Inflation models are tested against the CMB anisotropy measurements, specifically the spectral tilt $n_s^{}$ of the scalar power spectrum and the tensor-to-scalar ratio $r$.
The Planck 2018 release determines the $n_s^{}$ and $r$ at the pivot scale $k=0.05~\mathrm{Mpc}^{-1}$ to be~\cite{Planck:2018jri} 
\begin{align}
n_s^{}=0.9649\pm 0.0084,\quad r<0.063,~(95\% ~\mathrm{C.L.})\;. \label{eq:inflation_obs}
\end{align}
This release, along with the BICEP/Keck results~\cite{BICEP:2021xfz}, rules out more than $40\%$ of single-field slow-roll models. However, over one hundred models still survive~\cite{Martin:2024qnn}. 

Considering reheating physics can help alleviate or even break the degeneracy among inflationary models~\cite{Martin:2010kz,Dai:2014jja,Martin:2014nya,Cook:2015vqa}. Reheating is the transition period from the end of inflation to the thermal universe and is one of the least understood periods in cosmology. During the slow-roll phase, the inflaton remains at a potential position with a very small slope. Reheating occurs when the inflaton approaches its potential minimum and begins to oscillate around it. As the inflaton oscillates, it transfers its energy to other degrees of freedom, with frequent interactions among them heating the universe to a thermal state. Reheating concludes when the universe becomes radiation-dominated. Reheating physics cannot be directly probed. However, the cosmic evolution during reheating influences the moment the pivot scale re-enters the horizon, thereby affecting inflationary observations.

The origin of the baryon asymmetry of the universe (BAU) may also be connected to the early universe. The current value of the baryon asymmetry, as measured in the CMB, can be expressed in terms of the baryon-to-entropy ratio as~\cite{Planck:2018vyg}
\begin{align}
Y_\mathrm{B}^{}=\frac{n_\mathrm{B}^{}}{s}=\left(8.72\pm 0.08\right)\times 10^{-11}\;,\label{eq:BAU_obs}
\end{align}
where $n_\mathrm{B}^{}$ is the baryon number density and $s$ is the entropy density.
Among the various mechanisms generating the BAU dynamically, leptogenesis~\cite{Fukugita:1986hr} is of particular interest because it involves right-handed neutrinos (RHNs), which are necessary for explaining neutrino mass. In leptogenesis, the CP-violating and lepton-number-violating decays of the RHNs generate a lepton asymmetry, which is then converted into a baryon asymmetry through electroweak sphaleron processes. Non-thermal leptogenesis describes a scenario in which the RHNs are produced without a thermal distribution, meaning they are generated non-thermally.

Non-thermal leptogenesis naturally connects to inflation, providing a framework where the mechanisms of baryon asymmetry generation align with the dynamics of the early universe. It has been investigated in many inflation models~\cite{Kumekawa:1994gx,Asaka:1999yd,Lazarides:1999dm,Jeannerot:2001qu,Senoguz:2003hc,Dent:2003dn,Dent:2005gx,Endo:2006nj,Antusch:2010mv,Khalil:2012nd,Murayama:1992ua,Ellis:2003sq,Pallis:2011ps,Pallis:2011gr,Pallis:2012iw,Antusch:2018zvu,Panotopoulos:2021ttt,SravanKumar:2018tgk,Fukuyama:2005us,Baer:2008eq,Fukuyama:2010hh} and various neutrino models~\cite{Asaka:2002zu,Allahverdi:2002gz,Panotopoulos:2006wj,Senoguz:2007hu}. 
In ref.~\cite{Zhang:2023oyo}, we explore the intriguing possibility that the inflaton couples to RHNs, which subsequently decay to Standard Model (SM) particles to reheat the universe. This provides a minimal framework that connects inflation with the BAU. The non-thermal generation of RHNs is critical for linking the observational constraints of inflation to the BAU, as thermally produced RHNs have no memory of their initial states.

In this work, we continue our investigation of the minimal framework by considering the reheating constraints on inflation models in detail. Assuming that inflaton couples only directly to RHNs, we introduce the concept of  ``neutrino reheating", wherein the universe is heated by neutrino-related interactions. In ref.~\cite{Zhang:2023oyo}, the inflationary observables are analyzed under generic assumptions regarding reheating, considering $50$ to $60$ e-folds from the end of inflation to horizon exit. In this work, we conduct a consistent calculation of inflationary observables within the context of neutrino reheating. We will demonstrate that neutrino reheating establishes a connection between the BAU and inflationary observables, providing specific predictions that may help break the degeneracy of inflationary models. Our focus is on the perturbative decay of the inflaton to RHNs, as Pauli blocking renders parametric resonance inefficient~\cite{Croon:2019dfw}. RHNs being responsible for reheating is also investigated in \cite{Cosme:2024ndc} and in the context of gravitational neutrino reheating~\cite{Haque:2023zhb,Haque:2024zdq}.

This work is organized as follows: in section~\ref{sec:RH} we review the reheating parameters and introduce the $N_\mathrm{RH}^{}$-$N_k^{}$ relation; in section~\ref{sec:NRH} we introduce neutrino reheating from inflaton decay and calculate e-folds during neutrino reheating in two scenarios; in section~\ref{sec:examples}, we analyze four concrete inflationary models with neutrino reheating and explore corresponding parameter spaces; we conclude in section~\ref{sec:conclusion}.

\section{Reheating parameters and the $N_\mathrm{RH}^{}$-$N_k^{}$ relation}\label{sec:RH}
Three quantities are used to parameterize the unknown reheating dynamics: the effective equation of state parameter $\omega_{\mathrm{RH}}^{}$, the duration of reheating in terms of e-folds $N_\mathrm{RH}^{}$, and the reheating temperature $T_\mathrm{RH}^{}$, which marks the temperature at the end of reheating. During reheating, the energy density can be expressed as $\rho=\rho_\mathrm{end}^{}(a/a_\mathrm{end}^{})^{-3(1+\omega_{\mathrm{RH}}^{})}$, where $a$ is the scale factor and the subscripts $_\mathrm{end}^{}$ mark the end of inflation.
By definition, we have
\begin{align}
N_\mathrm{RH}^{} \equiv \mathrm{ln} \displaystyle \frac{a_\mathrm{RH}^{}}{a_\mathrm{end}^{}}=\frac{1}{3(1+\omega_\mathrm{RH}^{})} \mathrm{ln}\left(\frac{\rho_\mathrm{end}^{}}{\rho_\mathrm{RH}^{}}\right)\,, \label{eq:Nrh}
\end{align}
where in the last equality we use the energy density redshift relation. The energy density at the end of inflation is $\rho_\mathrm{end}^{}=3 V_\mathrm{end}^{}/2$ (with the equation of state parameter $\omega=-1/3$), and the energy density at the end of reheating is $\rho_\mathrm{RH}^{}=g_*^{} \pi^2 T_\mathrm{RH}^4/30$.
Since the reheating temperature is unknown in many scenarios, one can re-express $T_\mathrm{RH}^{}$ with $N_\mathrm{RH}^{}$ and other quantities. The entropy conservation after reheating gives
\begin{align}
g_{*s,\mathrm{RH}}^{} T_\mathrm{RH}^3 a_\mathrm{RH}^3=\left(2T_0^{3}+6\times \frac{7}{8} T_{\nu 0}^3\right) a_0^3 \,,
\end{align}
where $T_{\nu 0}^{}=(4/11)^{1/3}T_0^{}$. Then one finds
\begin{align}
T_\mathrm{RH}^{}=\left( \frac{43}{11g_{*s,\mathrm{RH}^{}}}\right)^{1/3} \frac{a_0^{}
}{a_\mathrm{RH}^{}} T_0^{} 
\end{align}
The ratio of the scale factors can be written as
\begin{align}
\frac{a_0^{}}{a_\mathrm{RH}^{}}=\frac{a_k^{}}{a_\mathrm{end}^{}} \frac{a_\mathrm{end}^{}}{a_\mathrm{RH}^{}} \frac{a_0^{}H_k^{}}{a_k^{}H_k^{}}=\frac{a_k^{}}{a_\mathrm{end}^{}} \frac{a_\mathrm{end}^{}}{a_\mathrm{RH}^{}} \frac{a_0^{}H_k^{}}{k}\,. \label{eq:a-ratio}
\end{align}
Then $T_\mathrm{RH}^{}$ can be expressed as
\begin{align}
T_\mathrm{RH}^{}=\left( \frac{43}{11g_{*s,\mathrm{RH}^{}}}\right)^{1/3} e^{-N_k^{}} e^{-N_\mathrm{RH}^{}} \frac{a_0^{}H_k^{}}{k} T_0^{} \,. \label{eq:Trh-T0}
\end{align}
Substituting eq.~(\ref{eq:Trh-T0}) in eq.~(\ref{eq:Nrh}), one finds the following relation
\begin{align}
N_\mathrm{RH}^{}=\frac{4}{3(1+\omega_\mathrm{RH}^{})} \left[ \mathrm{ln}\frac{V_\mathrm{end}^{1/4}}{H_k^{}} +\frac{1}{4}\mathrm{ln}\frac{45}{\pi^2g_*}-\frac{1}{3}\mathrm{ln} \frac{43}{11 g_{*s,\mathrm{RH}}} + \mathrm{ln} \frac{k}{a_0^{}T_0^{}} +N_k^{} +N_\mathrm{RH}^{}\right]\,.   \label{eq:Nk-Nrh}
\end{align}
One can solve for $N_\mathrm{RH}^{}$ if $\omega_\mathrm{RH}^{}\neq 1/3$ and find
\begin{align}
N_\mathrm{RH}^{}=\frac{4}{(1-3\omega_\mathrm{RH}^{})} \left[ -\mathrm{ln}\frac{V_\mathrm{end}^{1/4}}{H_k^{}} -\frac{1}{4}\mathrm{ln}\frac{45}{\pi^2g_*}+\frac{1}{3}\mathrm{ln} \frac{43}{11 g_{*s,\mathrm{RH}}} -\mathrm{ln} \frac{k}{a_0^{}T_0^{}} -N_k^{} \right]\,. \label{eq:Nk-Nrh2}
\end{align}
If $\omega_\mathrm{RH}= 1/3$, $N_\mathrm{RH}^{}$ cancels out from both sides of eq.~\eqref{eq:Nk-Nrh}, one arrives at \begin{align}
0=\mathrm{ln}\frac{V_\mathrm{end}^{1/4}}{H_k^{}} +\frac{1}{4}\mathrm{ln}\frac{45}{\pi^2g_*}-\frac{1}{3}\mathrm{ln} \frac{43}{11 g_{*s,\mathrm{RH}}} + \mathrm{ln} \frac{k}{a_0^{}T_0^{}} +N_k^{}\,. \label{eq:w=1/3}
\end{align}
Assuming $g_*\simeq g_{*s,\mathrm{RH}}\simeq 100$ and taking the pivot scale at $k=0.05 \mathrm{Mpc}^{-1}$, the above relation reduces to \cite{Cook:2015vqa}
\begin{align}
\mathrm{ln}\frac{V_\mathrm{end}^{1/4}}{H_k^{}}+N_k^{}=61.6\,.
\end{align}
Note that the same equation \eqref{eq:w=1/3} can be obtained if $N_\mathrm{RH}^{}=0$. $N_\mathrm{RH}^{}=0$ means there is no separate reheating phase. Actually, $\omega_\mathrm{RH}^{}= 1/3$ is difficult to conceive since it corresponds to the equation of state of a radiation-dominated phase.
The $N_\mathrm{RH}^{}$-$N_k^{}$ relation in \eqref{eq:Nk-Nrh} (and \eqref{eq:Nk-Nrh2}) explicitly shows that reheating dynamics affects inflationary observables
and is used to refine CMB constraints on inflation in refs.~\cite{Martin:2010kz,Dai:2014jja,Martin:2014nya,Cook:2015vqa,Drewes:2015coa,Ghoshal:2022fud,Chakraborty:2023ocr,German:2023yer,Ballardini:2024ado,German:2024rmn}.

There are three quantities related to inflationary models appearing in the $N_\mathrm{RH}^{}$-$N_k^{}$ relation, the potential energy at the end of inflation $V_\mathrm{end}^{}$, the e-folds from the end of inflation to the horizon exit $N_k^{}$ and the Hubble parameter at the horizon exit $H_k^{}$. Once we specify an inflationary model, these quantities can be expressed in terms of the model parameters and then connected to reheating dynamics via eq.~\eqref{eq:Nk-Nrh}.

\section{Neutrino reheating and non-thermal leptogenesis}\label{sec:NRH}
As the name suggests, by neutrino reheating, we consider that the universe is heated after inflation by neutrino-related interactions. We assume that the inflaton only couples directly to the RHNs. As inflation ends, the inflaton decays into the RHNs, which then decay into radiation. We consider this possibility for reheating because the RHNs naturally extend the SM to account for small neutrino masses and explain the BAU via leptogenesis. This scenario arises naturally from inflaton decay when assuming a direct inflaton-RHN coupling.

In ref.~\cite{Zhang:2023oyo}, neutrino reheating is studied implicitly when considering non-thermal leptogenesis from inflaton decay. The classification of non-thermal leptogenesis also serves as a classification of neutrino reheating. Depending on the relative strengths of the inflaton decay rate $\Gamma_\phi$ and the RHN decay rate $\Gamma_N$,
we can have: 1) $\Gamma_N^{}\geq \Gamma_\phi^{}$, RHNs almost decay instantly as they are produced, which we call ``instantaneous RHN decay"~\cite{Zhang:2023oyo}. 2) $\Gamma_N^{}< \Gamma_\phi^{}$, RHNs will survive for a while, leading to an RHN dominant phase if neutrino interactions are weak enough. It is the ``delayed RHN decay" scenario for non-thermal leptogenesis. These two classes cover the full parameter space, but we find analytic expressions for the final baryon asymmetry only in two limits.

In the $\Gamma_N^{}\gg \Gamma_\phi^{}$ limit, the RHNs decay instantly as they are produced. The final baryon asymmetry can be well approximated to be~\cite{Zhang:2023oyo}
\begin{align}
Y_\mathrm{B}^{}=\frac{3}{2}c_\mathrm{sph}^{}\epsilon \frac{T_\mathrm{RH}^{}}{M_\phi^{}}\,,\label{eq:YB1}
\end{align}
where $c_\mathrm{sph}^{}=28/79$ is the B$-$L to B conversion factor in sphaleron processes, $\epsilon$ is the RHN decay asymmetry.
Since the RHNs decay very quickly into radiation, overall, this scenario resembles the inflaton passing its energy directly to radiation, indicating $N_\mathrm{RH}^{}=0$. It can be inferred from the two lower evolution plots in figures.1 and 2 in ref.~\cite{Zhang:2023oyo}. As mentioned, $N_\mathrm{RH}=0$ also leads the $N_\mathrm{RH}^{}$-$N_k^{}$ relation to its reduced form~\eqref{eq:w=1/3}. Meanwhile, from eq.~(\ref{eq:Nrh}), $N_\mathrm{RH}^{}=0$ is equivalent to 
\begin{align}
\mathrm{ln}\left(\frac{3V_\mathrm{end}^{}}{2\rho_\mathrm{RH}^{}}\right)=0\,.\label{eq:Nrh0}
\end{align}
Since $\rho_\mathrm{RH}^{}=(\pi^2g_*/30)T_\mathrm{RH}^4$, one can use eq.~\eqref{eq:Nrh0} to express the reheating temperature $T_\mathrm{RH}^{}$ in terms of inflationary model parameters.

In the $\Gamma_N^{}\ll \Gamma_\phi^{}$ limit, the final baryon asymmetry can be written as~\cite{Zhang:2023oyo}
\begin{align}
Y_\mathrm{B}^{}=\frac{3}{4}c_\mathrm{sph}^{}\epsilon \frac{T_*^{}}{M_N^{}}\,, \label{eq:YB2}
\end{align}
where $T_*^{}$ is the RHN decay temperature. This limit features a delayed RHN decay and a RHN matter dominant phase. Once the RHNs decay efficiently, the universe becomes radiation-dominated. As a result, $T_*^{}$ is the reheating temperature in this scenario and we will relabel it as $T_\mathrm{RH}^{}$ from now on.

Since there exists an RHN matter-dominated phase, the reheating duration $N_\mathrm{RH}^{}$ is nonzero. It can be calculated as follows. We first consider the possibility that the RHNs are produced relativistically with energy $E_N^{}=M_\phi^{}/2$ at $T_\phi^{}$, where $T_\phi^{}$ marks the time the inflaton decays efficiently. As the universe cools down, the RHNs become non-relativistic at a temperature $T_\mathrm{NR}^{}=T_\phi^{}M_N^{}/E_N^{}$ and begin to dominate the universe as matter. At $T_\mathrm{RH}^{}$, RHNs decay efficiently to radiation, and the reheating ends. We can write the energy density at each temperature (time)\footnote{Note that $T_\mathrm{RH}^{}$ may not be the highest temperature during reheating~\cite{Chung:1998rq}. Additionally, temperature may not be well defined without a thermal bath. $T_\phi^{}$ and $T_\mathrm{NR}^{}$ here should be understood symbolically as a measure of the cosmic time previous to (the cosmic time corresponds to) $T_\mathrm{RH}$.}
\begin{align}
\rho_N^{}|_{T_\phi}&=\rho_\phi^{}|_{T_\phi}=E_N^{}n_N^{}\,, \label{eq:rhoTphi}\\
\rho_N^{}|_{T_\mathrm{NR}}&=M_N^{}n_N^{}\,,\label{eq:rhoTnr}\\
\rho_N^{}|_{T_\mathrm{RH}}&=\rho_\mathrm{R}^{}|_{T_\mathrm{RH}}\,,
\end{align}
where $\rho_N^{}|_{T_\phi}^{}$ is $\rho_N^{}$ at $T_\phi^{}$.
When the RHNs are relativistic, they redshift as radiation and feature an equation of state parameter $\omega_\mathrm{I}^{}=1/3$. When the RHNs become non-relativistic, they redshift as matter with an equation of state parameter $\omega_\mathrm{II}^{}=0$. Since the main constituent in both phases is the RHNs, we can calculate the e-folds in these two phases as follows
\begin{align}
N_\mathrm{I}^{}&=\frac{1}{3(1+\omega_\mathrm{I}^{})}\mathrm{ln} \frac{\rho_N^{}|_{T_\phi}^{}}{\rho_N^{}|_{T_\mathrm{NR}}^{}}=\mathrm{ln}\frac{M_\phi^{}}{2M_N^{}}\,.\label{eq:N1}\\
N_\mathrm{II}^{}&=\frac{1}{3(1+\omega_\mathrm{II}^{})}\mathrm{ln} \frac{\rho_N^{}|_{T_\mathrm{NR}}^{}}{\rho_N^{}|_{T_\mathrm{RH}}^{}} \nonumber\\
&=\frac{1}{3}\left(\mathrm{ln}\rho_\phi^{}|_{T_\phi^{}}\left(\frac{M_N^{}}{E_N^{}}\right)^4 - \mathrm{ln} \frac{\pi^2g_*^{}}{30}T_\mathrm{RH}^4\right) \nonumber\\
&=\frac{1}{3}\left(\mathrm{ln}\frac{720}{\pi^2g_*^{}} +\mathrm{ln}\frac{V_\mathrm{end}^{}}{M_\phi^4} +4\mathrm{ln}\frac{M_N^{}}{T_\mathrm{RH}^{}}
\right)\,, \label{eq:N2}
\end{align}
where we first translate $\rho_N^{}|_{T_\mathrm{NR}}^{}$ to $\rho_N^{}|_{T_\phi}^{}$, then to $\rho_\phi^{}|_{T_\phi}^{}$ using eq.~\eqref{eq:rhoTphi}. We also approximate $\rho_\phi^{}|_{T_\phi^{}}\simeq \rho_\phi^{}|_\mathrm{end}$ by assuming that the inflaton decays soon after the end of inflation. Now we can easily express the reheating e-folds as $N_\mathrm{RH}^{}=N_\mathrm{I}^{}+N_\mathrm{II}^{}$. From the expression of $N_\mathrm{RH}^{}$, we can deduce the averaged equation of state parameter $\omega_\mathrm{RH}^{}$ from eq.~\eqref{eq:Nrh}.

Now we are ready to discuss some constraints on the parameters in the scenario we considered. First, all the e-folds should be non-negative. $N_\mathrm{I}^{}$ in \eqref{eq:N1} naturally satisfies $N_\mathrm{I}^{}\geq 0$, while $N_\mathrm{II}^{}\geq 0$ sets an upper limit for $T_\mathrm{RH}^{}$
\begin{align}
T_\mathrm{RH}^{} \leq \left(\frac{30\rho_N^{}|_{T_\phi^{}}}{\pi^2 g_*^{}} \right)^{1/4}\frac{2M_N^{}}{M_\phi^{}}\,.\label{eq:Trhlimit}
\end{align}
A lower limit on $T_\mathrm{RH}^{}$ is set by the Big Bang nucleosynthesis (BBN) to be a few MeV~\cite{Kawasaki:2000en,Hasegawa:2019jsa,Iocco:2008va}.
Second, since $\omega_\mathrm{I}^{}=1/3$ and $\omega_\mathrm{II}^{}=0$, the average reheating equation of state parameter is expected to be between $0$ and $1/3$ and thus positive. As such, the $N_\mathrm{RH}^{}$-$N_k^{}$ relation in \eqref{eq:Nk-Nrh2} indicates
\begin{align}
-\mathrm{ln}\frac{V_\mathrm{end}^{1/4}}{H_k^{}} -\frac{1}{4}\mathrm{ln}\frac{45}{\pi^2g_*}+\frac{1}{3}\mathrm{ln} \frac{43}{11 g_{*s,\mathrm{RH}}} -\mathrm{ln} \frac{k}{a_0^{}T_0^{}} -N_k^{}>0\,,\label{eq:p-condition}
\end{align}
which gives an upper limit on $n_s^{}$. A lower limit on $n_s^{}$ can be determined by imposing a lower limit on $N_k^{}$, which ensures that inflation solves the flatness and horizon problems and that reheating occurs above the BBN scale~\cite{Cook:2015vqa}. The discussions above will help identify a reasonable parameter space once an inflation model is specified.

In an intermediate scenario where the analytical approximations for the final baryon asymmetry do not apply, one must solve the Boltzmann equations to understand the dynamics. From the numerical results of the Boltzmann equations, information can be obtained about the duration of the reheating phase $N_\mathrm{RH}^{}$. Combined with the energy density at the end of inflation and at the end of reheating, the effective equation of state parameter $\omega_\mathrm{RH}^{}$ can be deduced. With this information, predictions for neutrino reheating can be made using the $N_\mathrm{RH}^{}$-$N_k$ relation.

The discussions on neutrino reheating presented above are general and apply to any inflation model. To obtain specific predictions, one must work with a particular inflation model. We will present several examples in the next section.

\section{Neutrino reheating brings $Y_\mathrm{B}^{}$ and $n_s^{}$ together}\label{sec:examples}

In this section, we show examples in which, due to neutrino reheating, the BAU parameter can be expressed as a function of the spectral index, allowing very specific predictions in each model.

\subsection{Example 1: A polynomial type potential}\label{sec:poly}
We consider first the inflaton possesses a polynomial type potential of the following form
\begin{align}
V=\frac{1}{2}m^{4-\alpha}\phi^{\alpha}\,. \label{eq:Vpoly}
\end{align}
The effects of reheating in this potential are discussed in refs.~\cite{Martin:2013tda,Dai:2014jja,Martin:2014nya,Cook:2015vqa}.
Now we calculate the inflationary model-dependent quantities shown in the $N_\mathrm{RH}^{}$-$N_k^{}$ relation~\eqref{eq:Nk-Nrh}. The number of e-folds $N_k^{}$ is
\begin{align}
N_k^{}=\frac{1}{2\alpha} \frac{\phi_k^2-\phi_\mathrm{end}^2}{M_\mathrm{P}^2}\simeq \frac{1}{2\alpha} \frac{\phi_k^2}{M_\mathrm{P}^2} \,,
\end{align}
where $M_\mathrm{P}^{}$ is the reduced Planck mass. In the last equality, we omit the $\phi_\mathrm{end}^{}$ term by assuming that the potential is flat enough such that $\phi_k^{} \gg \phi_\mathrm{end}^{}$. This relation allows us to express the slow-roll parameters (hence the spectral index and the tensor-to-scalar ratio) in terms of $N_k^{}$.
\begin{align}
\epsilon_k^{} &=\frac{\alpha^2}{2}\frac{M_\mathrm{P}^2}{\phi_k^2}=\frac{\alpha}{4N_k^{}} \label{eq:ek}\\
\eta_k^{}&=\alpha(\alpha-1)\frac{M_\mathrm{P}^2}{\phi_k^2}=\frac{\alpha-1}{2N_k^{}}\\
n_s^{}&=1-6\epsilon_k^{}+2\eta_k^{}=1-\frac{\alpha+2}{2N_k^{}}\label{eq:poly-ns}\\
r&=16\epsilon_k^{}=\frac{4\alpha}{N_k^{}}\,.\label{eq:poly-r}
\end{align}
Eq.~\eqref{eq:poly-ns} allows us to express $N_k^{}$ in terms of $n_s^{}$
\begin{align}
N_k^{}=\frac{\alpha+2}{2(1-n_s^{})}\,.\label{eq:poly-Nk}
\end{align}
Using $\epsilon_k^{}$ in \eqref{eq:ek} and the relation in \eqref{eq:poly-Nk}, $H_k^{}$ can be expressed as
\begin{align}
H_k^{}=\pi M_\mathrm{P}^{} \sqrt{8 A_s^{}\epsilon_k^{}}
=2\pi M_\mathrm{P}^{} \sqrt{\frac{\alpha}{\alpha+2}A_s^{}(1-n_s^{})}\,.\label{eq:poly-Hk}
\end{align}
It then allows us to write $V_\mathrm{end}$ as
\begin{align}
V_\mathrm{end}^{}=V_k^{}\left(\frac{\phi_\mathrm{end}^{}}{\phi_k^{}}\right)^\alpha =3M_\mathrm{P}^2H_k^2\left(\frac{\phi_\mathrm{end}^{}}{\phi_k^{}}\right)^\alpha
=12\pi^2M_\mathrm{P}^4 \frac{\alpha}{\alpha+2}A_s^{}(1-n_s^{})\left(\frac{\alpha(1-n_s^{})}{2(\alpha+2)}\right)^{\alpha/2}\,.
\end{align}
Now, the three model-dependent quantities shown in the $N_\mathrm{RH}^{}$-$N_k^{}$ relation in eq.~\eqref{eq:Nk-Nrh} are all expressed in terms of the model parameter $\alpha$ and the spectral index $n_s$.

We need a non-zero inflaton mass to allow decaying to RHNs. Since $M_\phi^2=V^{''}|_\mathrm{min}$, it is nonzero only for $\alpha=2$ and we will focus on this case from now on. In this case, we have $M_\phi=m\simeq 8\times 10^{13}$ GeV, where the model parameter $m$ is calculated by matching the potential \eqref{eq:Vpoly} at horizon exit with observations using the slow-roll approximation $V_k^{}\simeq 3M_\mathrm{P}^2H_k^2$ with $H_k^{}$ in eq.~\eqref{eq:poly-Hk}.

\vspace{0.2cm}
\noindent {\bf Neutrino reheating predictions in the $\Gamma_N^{}\gg \Gamma_\phi^{}$ scenario}
\vspace{0.15cm}

We first consider neutrino reheating in the instantaneous RHN decay limit, that is, $\Gamma_N^{}\gg \Gamma_\phi^{}$. Since $N_\mathrm{RH}=0$, we use the reduced $N_\mathrm{RH}^{}$-$N_k^{}$ relation~\eqref{eq:w=1/3} to find a prediction of $n_s^{}=0.9652$.
$N_\mathrm{RH}^{}=0$ also leads to eq.~\eqref{eq:Nrh0}. Since $\rho_\mathrm{RH}^{}=(\pi^2g_*/30)T_\mathrm{RH}^4$, one can express the reheating temperature $T_\mathrm{RH}^{}$ as a function of $n_s^{}$ with relation determined by eq.~\eqref{eq:Nrh0}. The predicted $n_s^{}$ corresponds to $T_\mathrm{RH}^{}=2.79\times 10^{15}$ GeV. The final baryon asymmetry can be estimated from eq.~\eqref{eq:YB1} as\footnote{We use $\epsilon=10^{-6}$ as a reasonable approximation of the RHN decay asymmetry.} $Y_\mathrm{B}^{}=1.85\times 10^{-5}$, several orders of magnitude larger than observed. 
The tensor-to-scalar ratio in this model is a function of $\alpha$ and $n_s^{}$~\eqref{eq:poly-r}. With $\alpha=2$ and the predicted $n_s^{}$, we find $r=0.139$, which is not allowed given the latest constraints~\cite{BICEP:2021xfz}.

\vspace{0.2cm}
\noindent{\bf Neutrino reheating predictions in the $\Gamma_N^{}\ll \Gamma_\phi^{}$ scenario}
\vspace{0.15cm}

The neutrino reheating duration $N_\mathrm{RH}^{}$ calculated from \eqref{eq:N1} and \eqref{eq:N2} depends on the following four parameters (since nonzero inflaton mass fixes $\alpha=2$): $M_\phi^{},~M_N^{}$, $M_N^{}$, $T_\mathrm{RH}^{}$ and $n_s^{}$. Since the inflaton mass is calculated to be $M_\phi^{}\simeq8\times 10^{13}$ GeV, we are left with three variables 
$$M_N^{},~T_\mathrm{RH}^{},~n_s^{}\,.$$

\begin{figure}[t!]
\centering
\includegraphics[width=\textwidth]{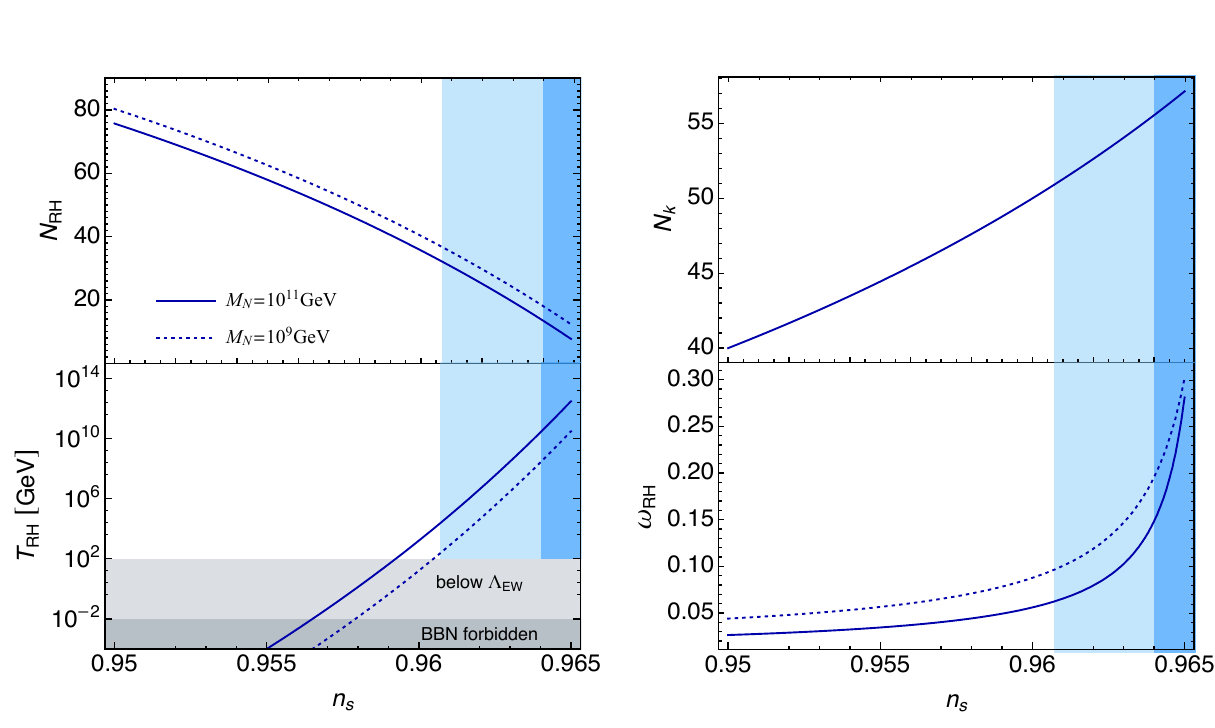}
\caption{The e-folds during reheating $N_\mathrm{RH}^{}$, the reheating temperature $T_\mathrm{RH}^{}$, the e-folds from the end of inflation to the exit of horizon $N_k^{}$, the effective equation of state parameter during reheating $\omega_\mathrm{RH}^{}$ in polynomial type inflation model ($\alpha=2$) with neutrino reheating in $\Gamma_N^{}\ll \Gamma_\phi^{}$ scenario. The dark gray shaded region in the first column marks the BBN forbidden region, and the light gray region is below the electroweak scale. The vertical light blue bands show allowed $1\sigma$ region of $n_s^{}$ given by Planck 2018~\cite{Planck:2018jri}. The blue band marks $1\sigma$ region of a future CMB experiment~\cite{Amendola:2016saw,PRISM:2013ybg}.}
\label{fig:poly1}
\end{figure}

\begin{figure}[t!]
\centering
\includegraphics[width=\textwidth]{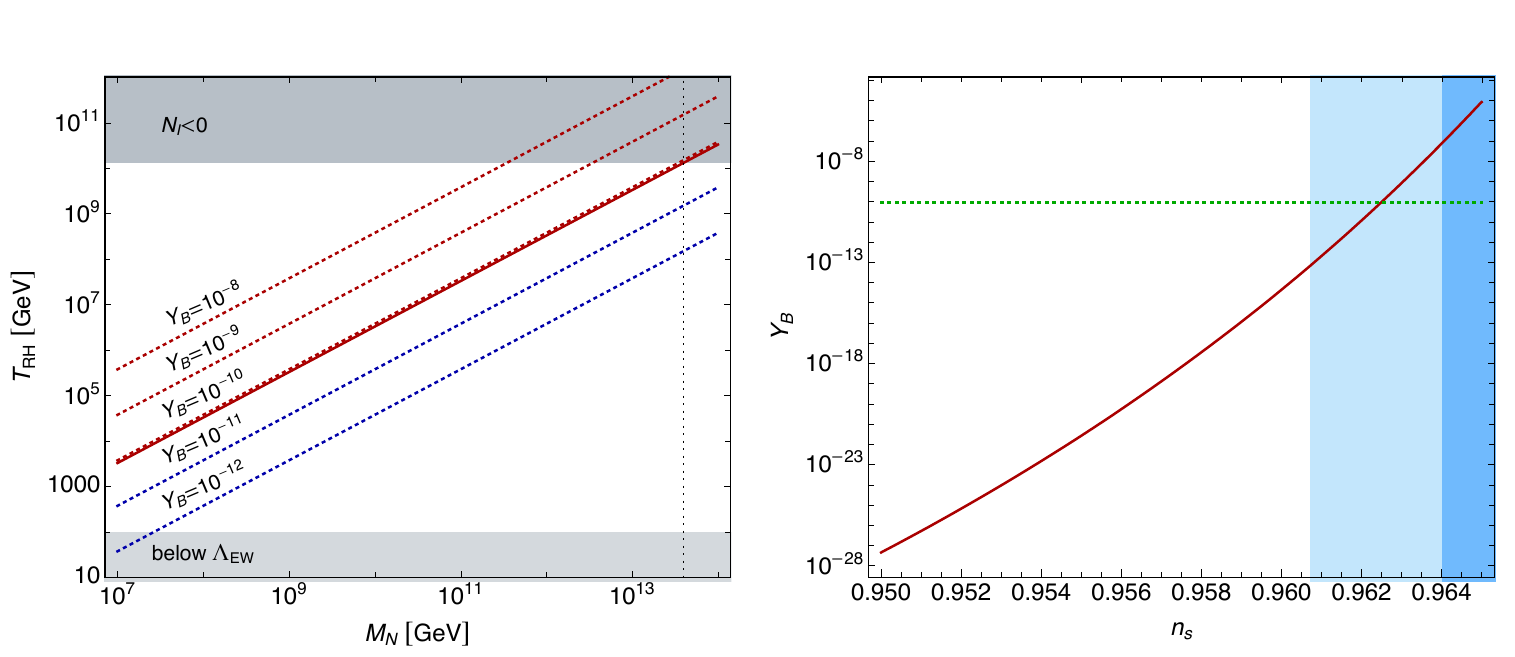}
\caption{The full parameter space in polynomial type inflation model ($\alpha=2$) with neutrino reheating in the $\Gamma_N^{}\ll \Gamma_\phi^{}$ scenario. In the left panel, the viable parameter space corresponds to the upper-left region above the red line. The dark gray shaded area at the top is excluded as it results in a negative $N_\mathrm{I}^{}$. The vertical dashed line indicates the upper bound on $M_N^{}$, derived from $M_\phi^{}/2$. The lower limit on $M_N^{}$ is adopted from \cite{Zhang:2023oyo}. In the right panel, the red curve shows the predicted $Y_\mathrm{B}^{}$ as a function of $n_s^{}$, while the green dotted line marks the observed baryon asymmetry value. The remaining shaded regions follow the same conventions as in the previous figure.}
\label{fig:poly2}
\end{figure}

\begin{figure}[t!]
\centering
\includegraphics[width=0.6\textwidth]{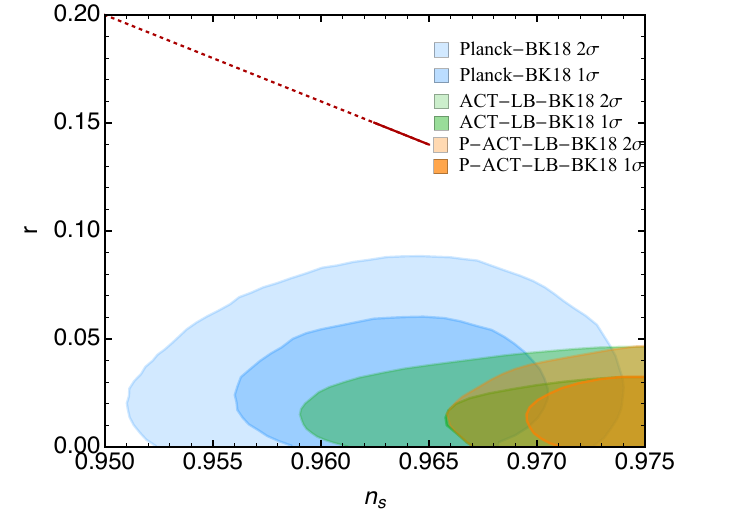}
\caption{Predictions for the polynomial type inflation model ($\alpha=2$) with neutrino reheating in the $\Gamma_N^{}\ll \Gamma_\phi^{}$ scenario. The red dotted line shows the predicted tensor-to-scalar ratio, with the continuous part corresponds to the range of $n_s^{}$ values that yield a sufficiently large baryon asymmetry $Y_\mathrm{B}^{}$. The Planck-BK18 contours adopted from the latest update of Ref.\cite{Martin:2013tda}, with detailed derivations provided in \cite{Martin:2024qnn}, incorporating both Planck and BICEP/Keck data. The ACT-LB-BK18 contour is based on \cite{ACT:2025tim}, using Atacama Cosmology Telescope (ACT) DR6 data combined with CMB lensing and baryon acoustic oscillation (BAO) measurements. The P-ACT-LB-BK18 contour additionally includes Planck data~\cite{ACT:2025tim}.}
\label{fig:poly3}
\end{figure}

We choose two benchmark values of the RHN mass to show parameter correlations for interested quantities. The $N_\mathrm{RH}^{}$-$N_k^{}$ relation allows to determine the parameter space in terms of $T_\mathrm{RH}^{}$ and $n_s^{}$. Using eq.~\eqref{eq:YB2}, we can establish a relationship between the baryon asymmetry and the spectral index.
The results are shown in figure~\ref{fig:poly1}, where we present various quantities as functions of the spectral index calculated from the $N_\mathrm{RH}^{}$-$N_k^{}$ relation with $N_\mathrm{RH}^{}$ calculated in the neutrino reheating scenario $\Gamma_N^{}\ll \Gamma_\phi^{}$. There is only one line in $N_k$ since it is only a function of $n_s^{}$ without dependence on $M_N$ \eqref{eq:poly-Nk}. In figure~\ref{fig:poly1}, a larger $M_N^{}$ corresponds to a smaller $N_\mathrm{RH}^{}$, $\omega_\mathrm{RH}^{}$ and a higher $T_\mathrm{RH}^{}$. $T_\mathrm{RH}$ is quite sensitive to $n_s^{}$, spanning several orders of magnitude, which is also observed in ref.~\cite{Cook:2015vqa}. As the lines approach $n_s^{}=0.9652$, $N_\mathrm{RH}^{}$ approaches zero.

To explore the full parameter space, we notice that both $M_N^{}$ and $T_\mathrm{RH}^{}$ show up proportionally in the expression for $Y_\mathrm{B}^{}$ in \eqref{eq:YB1} and in $N_\mathrm{RH}^{}$ via \eqref{eq:N2}. The observed $Y_\mathrm{B}$ constrains only the ratio of $T_\mathrm{RH}^{}/M_N^{}$. In the upper-left region of the $T_\mathrm{RH}^{}$-$M_N^{}$ plot in the left panel of figure~\ref{fig:poly2} (above the red line), the parameter space can produce a sufficiently large baryon asymmetry. Regarding the spectral index dependence, the plot in the right panel of figure~\ref{fig:poly2} shows that a viable $Y_\mathrm{B}^{}$ favors $n_s^{}>0.9625$. Additionally, $Y_\mathrm{B}^{}$ increases sharply with growing $n_s^{}$, spanning several tens of orders of magnitude. The predicted tensor-to-scalar ratio is shown in figure~\ref{fig:poly3}, where the $n_s^{}$ values preferred by successful baryogenesis are highlighted by the solid line. These results demonstrate that incorporating neutrino reheating—along with the requirement of sufficient baryon asymmetry—further restricts the parameter space.

\subsection{Example 2: Starobinsky model}\label{sec:star}
We consider the effective potential of Starobinsky inflation~\cite{Starobinsky:1980te} in the Einstein frame
\begin{align}
V=\Lambda^4 \left(1-e^{-\sqrt{\frac{2}{3}}\frac{\phi}{M_\mathrm{P}^{}}}\right)^2 \,.
\end{align}
The e-folds number from the end of inflation to horizon exit is
\begin{align}
N_k^{} =\frac{1}{M_\mathrm{P}^2} \int^{\phi_k^{}}_{\phi_\mathrm{end}^{}} \frac{M_\mathrm{P}^{}}{2}\sqrt{\frac{3}{2}}\left(e^{\sqrt{\frac{2}{3}}\frac{\phi}{M_\mathrm{P}^{}}}-1\right) \mathrm{d}\phi 
\simeq \frac{3}{4}e^{\sqrt{\frac{2}{3}}\frac{\phi_k^{}}{M_\mathrm{P}^{}}}\,
\end{align}
where in the last step we use approximation $\phi_k^{}\gg \phi_\mathrm{end}^{}$. It allows us to express $\phi_k^{}$ in terms of $N_k^{}$ as
\begin{align}
\phi_k^{}=\sqrt{\frac{3}{2}}M_\mathrm{P}^{} \mathrm{ln}\left(\frac{4}{3}N_k^{}\right)\,,
\end{align}
leading to the following neat expressions for the slow-roll parameters
\begin{align}
\epsilon_k^{}&=\frac{4}{3}\left(\frac{3}{4N_k^{}-3}\right)^2\simeq \frac{3}{4N_k^2}\,,\label{eq:star-epsilon}\\
\eta_k^{}&=\frac{3/2-N_k^{}}{(N_k^{}-3/4)^2}\simeq -\frac{1}{N_k^{}}\,.\label{eq:star-eta}
\end{align}
Similarly, the spectral index and the tensor-to-scalar ratio can be expressed in terms of $N_k^{}$ as
\begin{align}
n_s^{}&\simeq 1-\frac{2}{N_k^{}}\,,\label{eq:star-ns}\\
r&=\frac{12}{N_k^2}\,.\label{eq:star-r}
\end{align}
Eq.~\eqref{eq:star-ns} allows us to write 
\begin{align}
N_k^{}=\frac{2}{1-n_s^{}}\,,\label{eq:star-NKns}
\end{align}
substituting it back to eqs.\eqref{eq:star-epsilon} \eqref{eq:star-eta} \eqref{eq:star-r} allow us to express all these quantities in terms of $n_s^{}$.
With the relation in eq.~\eqref{eq:star-NKns}, the other interested quantities can be expressed as a function of $n_s^{}$ as
\begin{align}
H_k^{}&=\pi M_\mathrm{P}^{} \sqrt{6A_s^{}}\frac{1-n_s^{}}{2}\,,\\
V_\mathrm{end}^{}&= \frac{9}{2}\pi^2 M_\mathrm{P}^4 A_s^{}(1-n_s^{})^2 \left(\frac{\frac{1}{\frac{\sqrt{3}}{2}+1}}{1-\frac{3}{8}(1-n_s^{})}\right)^2\,.
\end{align}
The inflaton mass is 
\begin{align}
M_\phi^2=\frac{\mathrm{d}^2V}{\mathrm{d}\phi^2}|_\mathrm{min}^{}=\frac{4}{3}\frac{\Lambda^4}{M_\mathrm{P}^2}\,,
\end{align}
where $\Lambda\simeq 8\times 10^{15}$ GeV is fixed by the amplitude of the CMB anisotropies, yielding $M_\phi^{}=3\times 10^{13}$ GeV.

\vspace{0.2cm}
\noindent{\bf Neutrino reheating predictions in the $\Gamma_N^{}\gg \Gamma_\phi^{}$ scenario}
\vspace{0.15cm}

With the $N_\mathrm{RH}^{}$-$N_k^{}$ relation in \eqref{eq:w=1/3}, we find a prediction of $n_s^{}=0.964$, within the current $1\sigma$ region. The predicted $n_s$ corresponds to $T_\mathrm{RH}^{}=2.75\times 10^{15}$ GeV, leading to $Y_\mathrm{B}^{}=4.67\times 10^{-5}$, several orders larger than the observed value. It could be ``corrected" by considering particular suppressed values of the CP symmetry.

\vspace{0.2cm}
\noindent{\bf Neutrino reheating predictions in the $\Gamma_N^{}\ll \Gamma_\phi^{}$ scenario}
\vspace{0.15cm}

In this scenario, $N_\mathrm{RH}^{}$ calculated from \eqref{eq:N1} and \eqref{eq:N2} depends on four parameters: $M_\phi^{}$, $M_N^{}$, $T_\mathrm{RH}^{}$, $n_s^{}$. Since $M_\phi^{}=3\times 10^{13}$ GeV, we are left with three variables:
$$M_N^{},~T_\mathrm{RH}^{},~n_s^{}\,.$$ 
From \eqref{eq:p-condition} we can find an upper limit for $n_s^{}$. We find a lower limit for $n_s^{}$ from the requirement $N_k^{}\geq 29$ following the arguments in~\cite{Cook:2015vqa}. 

\begin{figure}[t!]
\centering
\includegraphics[width=\textwidth]{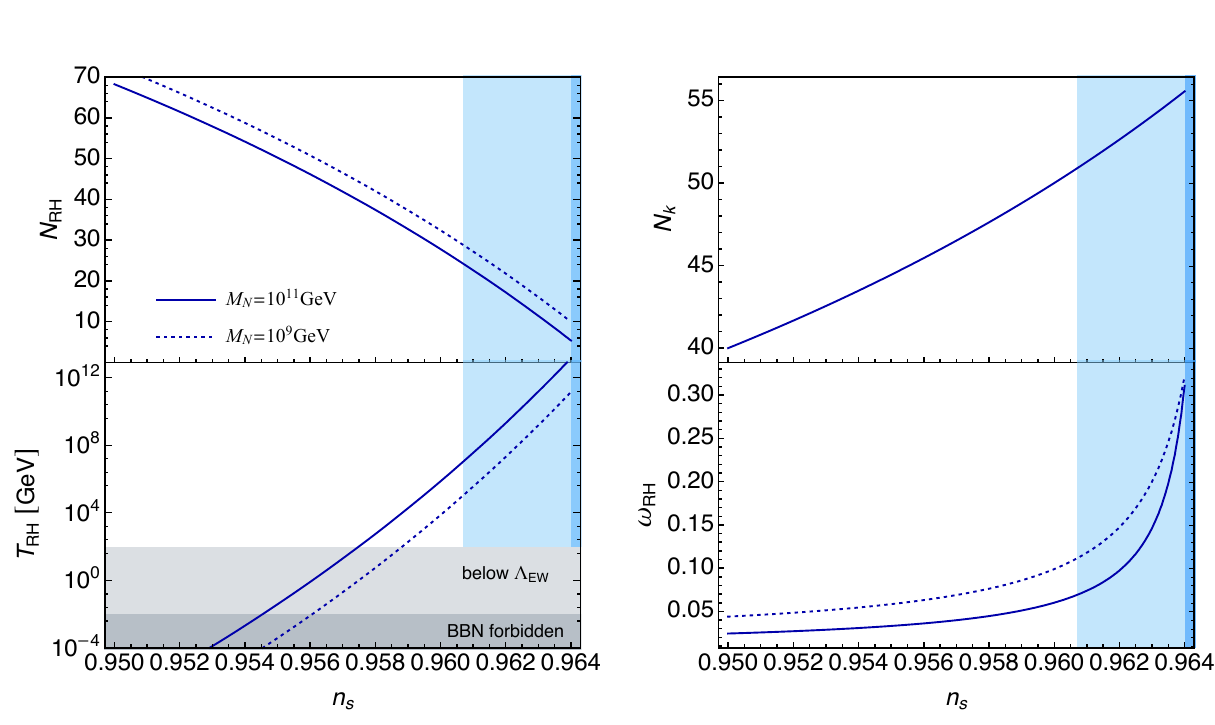}
\caption{Predictions in Starobinsky model with neutrino reheating in the $\Gamma_N^{}\ll \Gamma_\phi^{}$ scenario. The solid line corresponds to $M_N^{}=10^{11}$ GeV and the dotted line is $M_N^{}=10^{9}$ GeV. Other shadings are the same as in figure~\ref{fig:poly1}. }
\label{fig:star1}
\end{figure}

\begin{figure}[t!]
\centering
\includegraphics[width=\textwidth]{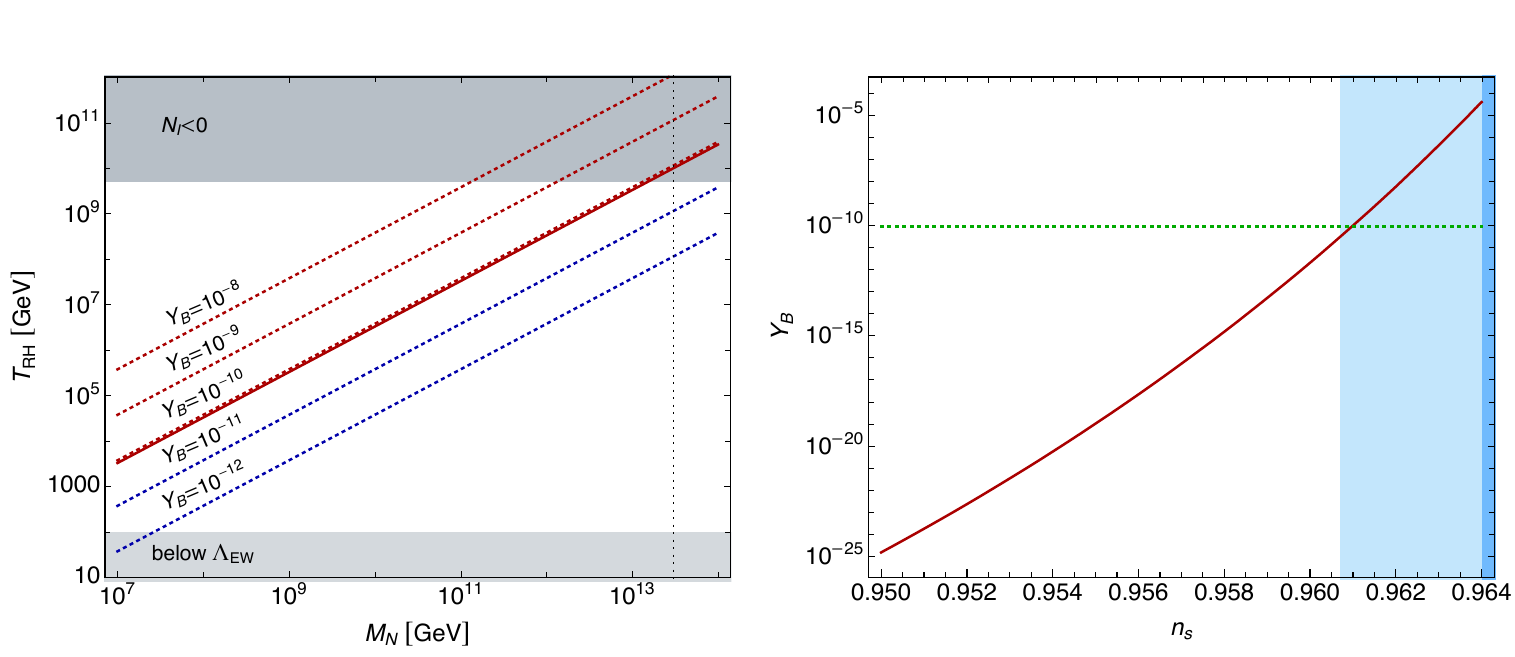}
\caption{The full parameter space in Starobinsky model with neutrino reheating in the $\Gamma_N^{}\ll \Gamma_\phi^{}$ scenario. The shaded regions follow the same conventions as in figure~\ref{fig:poly2}.}
\label{fig:star2}
\end{figure}

\begin{figure}[t!]
\centering
\includegraphics[width=0.6\textwidth]{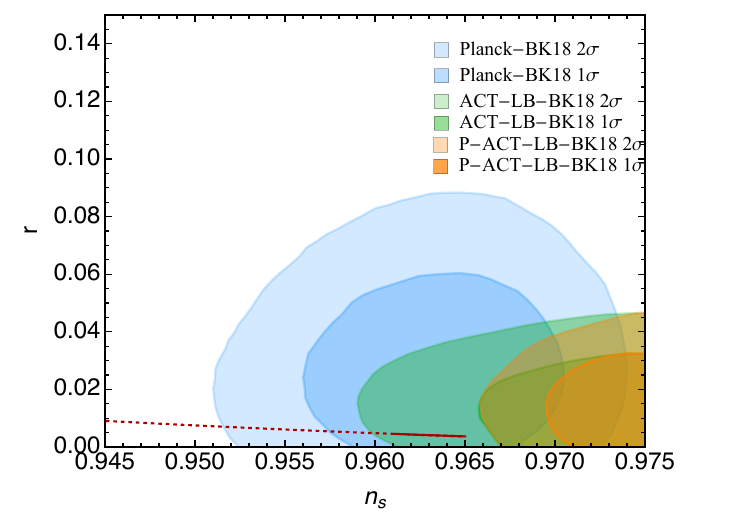}
\caption{Predictions for the Starobinsky model with neutrino reheating in the $\Gamma_N^{}\ll \Gamma_\phi^{}$ scenario. The red dotted line shows the predicted tensor-to-scalar ratio, while the solid portion corresponds to the range of $n_s^{}$ values that yield a sufficiently large baryon asymmetry. Shadings are the same as in figure~\ref{fig:poly3}.}
\label{fig:star3}
\end{figure}
We first choose to work with $M_N^{}=10^{11},10^9$ GeV. Then the $N_\mathrm{RH}^{}$-$N_k^{}$ relation allows us to determine $T_\mathrm{RH}^{}$ versus $n_s^{}$.
The numerical results are shown in figure~\ref{fig:star1}, where various quantities are displayed as functions of the spectral index. As in the polynomial potential, we find that a larger $M_N^{}$ corresponds to a smaller $N_\mathrm{RH}^{}$, a smaller $\omega_\mathrm{RH}^{}$, and a higher $T_\mathrm{RH}^{}$. Planck $1\sigma$ region of $n_s^{}$ corresponds to $N_\mathrm{RH}$ up to $28$ and $T_\mathrm{RH}$ in $[10^7,10^{13}]$ GeV.

The full parameter space is shown in figures~\ref{fig:star2} and \ref{fig:star3}. The left panel of figure~\ref{fig:star2} differs from that of figure~\ref{fig:poly2} only in the upper exclusion region, due to different inflaton masses.
In the right panel of figure~\ref{fig:star2}, we observe that $Y_\mathrm{B}^{}$ changes rapidly with $n_s^{}$. A sufficiently large enough $Y_\mathrm{B}^{}$ favors $n_s^{}>0.961$. When projected onto the $n_s^{}$-$r$ plane in figure~\ref{fig:star3}, it becomes evident that including $Y_\mathrm{B}^{}$ further constrains the parameter space (reducing the viable region from the entire dotted line to just the solid portion). Although $Y_\mathrm{B}^{}$ prefers larger $n_s^{}$ values - consistent with the latest ACT data release - achieving full agreement with observations remains challenging. The upper limit on $n_s^{}$ prevents exploration of higher $n_s^{}$ region.

The Starobinsky model in the Jordan frame adds the $R^2$ term to the Einstein-Hilbert action ($R$ is the Ricci scalar). After conformal transforming to the Einstein frame, it leads to inflaton coupling to all matter fields due to universal gravitational interactions, which is dubbed ``universal reheating"~\cite{Ketov:2025nkr} and is discussed in explaining BAU through leptogenesis in refs.~\cite{Gorbunov:2010bn,Jeong:2023zrv}.

\subsection{Example 3: Coleman-Weinberg inflation}\label{sec:cwi}
We consider a Coleman-Weinberg inflation expressed as~\cite{Coleman:1973jx,Rehman:2008qs,Okada:2014lxa}
\begin{align}
V=A \phi^4\left[\mathrm{ln}\left(\frac{\phi}{v}\right)-\frac{1}{4}\right]+\frac{1}{4}Av^4\,
\end{align}
where $v$ is the inflaton vacuum expectation value at the minimum, $V(0)\equiv V_0^{}=Av^4/4$ is the vacuum energy at origin. The amplitude of the scalar perturbation allows us to fix $V_0^{}$ for given $v$. The slow-roll parameters are
\begin{align}
\epsilon&=\displaystyle\frac{M_\mathrm{P}^2}{2}\left(\frac{16\phi^3 \mathrm{ln}\left(\frac{\phi}{v}\right)}{v^4-\phi^4+4\phi^4\mathrm{ln}\left(\frac{\phi}{v}\right)}\right)^2\,,\label{eq:epsilon}\\
\eta&=\displaystyle16M_\mathrm{P}^2 \frac{3\phi^2\mathrm{ln}\left(\frac{\phi}{v}\right)+\phi^2}{4\phi^4\mathrm{ln}\left(\frac{\phi}{v}\right)-\phi^4+v^4}\,.\label{eq:eta}
\end{align}
The e-folds number from the end of inflation to horizon exit is
\begin{align}
N_k^{}=\frac{1}{M_P^2}\left[\frac{1}{16} v^2 \text{Ei}\left(-2 \mathrm{ln} \left(\frac{\phi }{v}\right)\right)-\frac{1}{16} v^2 \text{Ei}\left(2 \mathrm{ln} \left(\frac{\phi }{v}\right)\right)+\frac{\phi ^2}{8}\right]|_{\phi_\mathrm{end}^{}}^{\phi_k^{}}\,,\label{eq:Nk-cw}
\end{align}
where Ei is the exponential integral function. Although the expressions may be somewhat lengthy, we can still proceed with our procedure numerically. With $\phi_\mathrm{end}^{}$ calculated from eq.~\eqref{eq:epsilon}, we can determine $\phi_k^{}$ as a function of $N_k^{}$ with eq.~\eqref{eq:Nk-cw}. Hence, the functions of $\phi_k$ are all expressed as functions of $N_k^{}$. We find $N_k(n_s)$ by converting $n_s^{}(N_k^{})$. This facilitates us to express $H_k^{}=\pi M_\mathrm{P}\sqrt{8A_s^{}\epsilon_k^{}}$ as $H_k^{}(n_s^{})$. For $V_\mathrm{end}^{}$, we use its relation with $V_k^{}$ and $H_k^{}$ to express it as $V_\mathrm{end}^{}(n_s^{})$.


\vspace{0.2cm}
\noindent{\bf Neutrino reheating predictions in the $\Gamma_N^{}\gg \Gamma_\phi^{}$ scenario}
\vspace{0.15cm}

\begin{figure}[t!]
\centering
\includegraphics[width=0.6\textwidth]{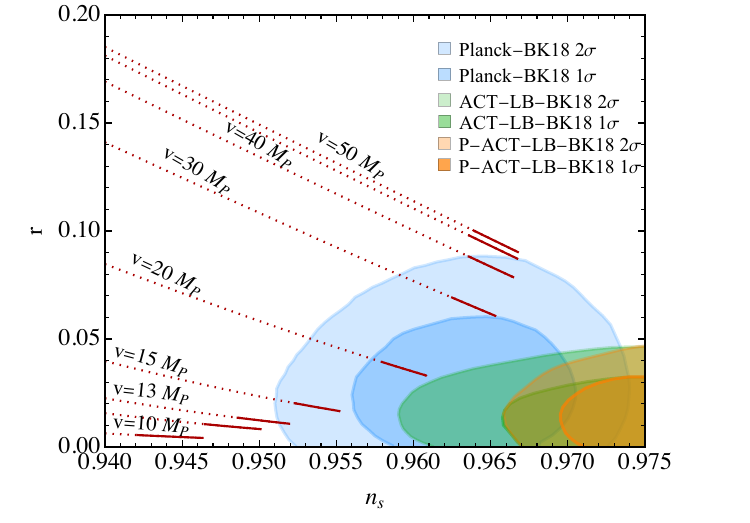}
\caption{Predictions in Coleman-Weinberg inflation with neutrino reheating in the $\Gamma_N^{}\ll \Gamma_\phi^{}$ scenario. The red dotted lines are the predicted tensor-to-scalar ratio. The continuous parts of the red lines indicate $n_s^{}$ values generating sufficiently large $Y_\mathrm{B}^{}$. Shadings are the same as in figure~\ref{fig:poly3}.}
\label{fig:cw3}
\end{figure}

In this scenario $N_\mathrm{RH}^{}=0$. Following the $N_\mathrm{RH}^{}$-$N_k^{}$ relation in \eqref{eq:w=1/3}, we find a prediction of $n_s^{}$ for given $v$. $N_\mathrm{RH}^{}=0$ also allows us to equate $\rho_\mathrm{end}^{}$ with $\rho_\mathrm{RH}^{}$. Hence, we find a relation of $T_\mathrm{RH}^{}$ with $n_s^{}$. The predictions correspond to the ending points in figure~\ref{fig:cw3}. The intersection with the Planck-BK18 $2\sigma$ region leads to the following estimation of the parameter space:
\begin{align}
&v \in [13,47] M_\mathrm{P}^{} \nonumber \\
&T_\mathrm{RH}^{}\in [2.47,2.91]\times 10^{15} \mathrm{GeV} \nonumber\\
&M_\phi^{} \in [1.40,1.71]\times 10^{13} \mathrm{GeV}\,.\nonumber
\end{align}
The predicted baryon asymmetry are from $9.05\times 10^{-5}$ to $9.39\times 10^{-5}$, several orders larger than needed, leaving much space for model building. 

\vspace{0.2cm}
\noindent{\bf Neutrino reheating predictions in the $\Gamma_N^{}\ll \Gamma_\phi^{}$ scenario}
\vspace{0.15cm}

In this scenario, the neutrino reheating duration $N_\mathrm{RH}^{}$, calculated from \eqref{eq:N1} and \eqref{eq:N2}, depends on the five parameters: $M_\phi^{}$,~$M_N^{}$,~$T_\mathrm{RH}^{}$,~$n_s^{}$,~$v$. The amplitude of the scalar perturbation allows us to determine $V_0^{}$ (and consequently $M_\phi^{}$) for a given $v$. We therefore select the following four independent parameters: 
$$M_N^{},~T_\mathrm{RH}^{},~v,~n_s^{}\,.$$

\begin{figure}[t!]
\centering
\includegraphics[width=\textwidth]{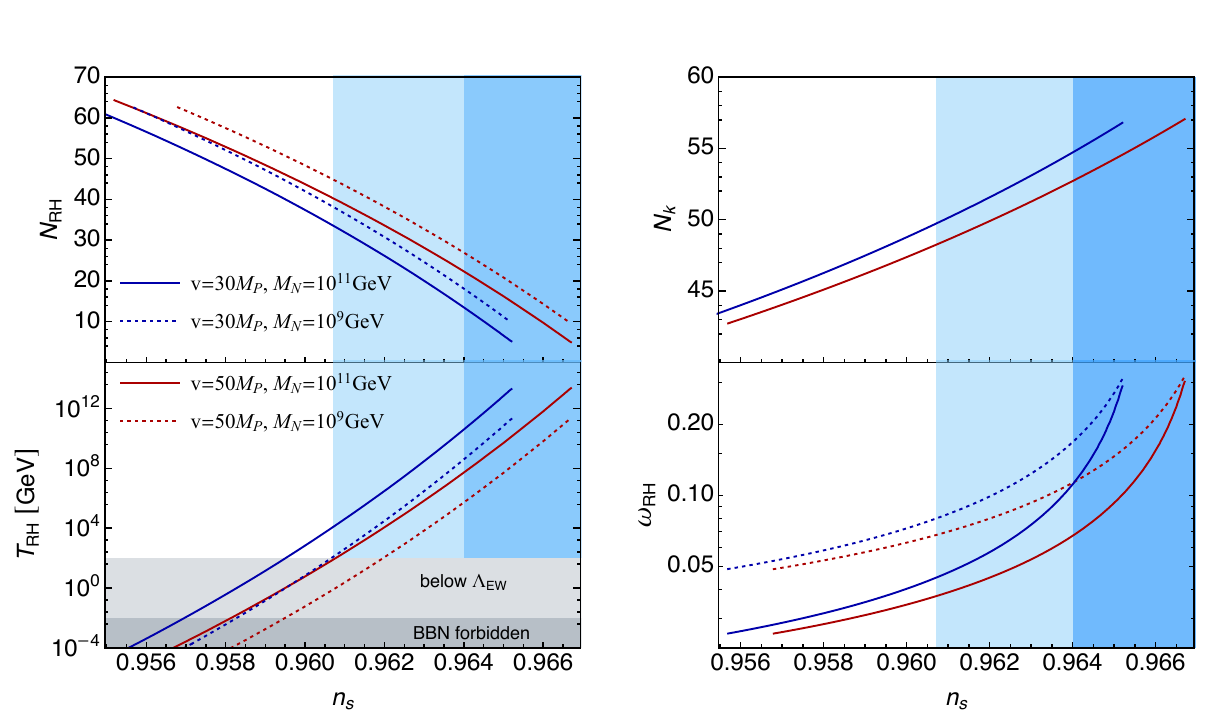}
\caption{Predictions in Coleman-Weinberg inflation with neutrino reheating in the $\Gamma_N^{}\ll \Gamma_\phi^{}$ scenario. Other shadings are the same as in the figure~\ref{fig:poly1}. }
\label{fig:cw1}
\end{figure}

To illustrate the correlations among the relevant quantities, we first consider benchmark values of $v=30 M_\mathrm{P}^{}, 50 M_\mathrm{P}^{}$ and $M_N=10^{11}, 10^9$ GeV. Using the $N_\mathrm{RH}^{}$-$N_k^{}$ relation, we then determine $T_\mathrm{RH}^{}$ as a function of $n_s^{}$. The results are shown in figure~\ref{fig:cw1}, where various quantities are shown as functions of the spectral index. In figure~\ref{fig:cw1}, we see that all lines overlap some with the observational region. Increasing the parameter $v$ leads to an extended reheating duration $N_\mathrm{RH}^{}$, a reduced reheating temperature $T_\mathrm{RH}^{}$, and smaller values of both $N_k^{}$ and $\omega_\mathrm{RH}^{}$. $T_\mathrm{RH}$ grows with $n_s^{}$ rapidly, spanning several orders of magnitude. For a fixed $v$, a larger $M_N$ corresponds to a higher $T_\mathrm{RH}^{}$, as well as smaller values of $N_\mathrm{RH}^{}$ and $\omega_\mathrm{RH}^{}$.


\begin{figure}[t!]
\centering
\includegraphics[width=\textwidth]{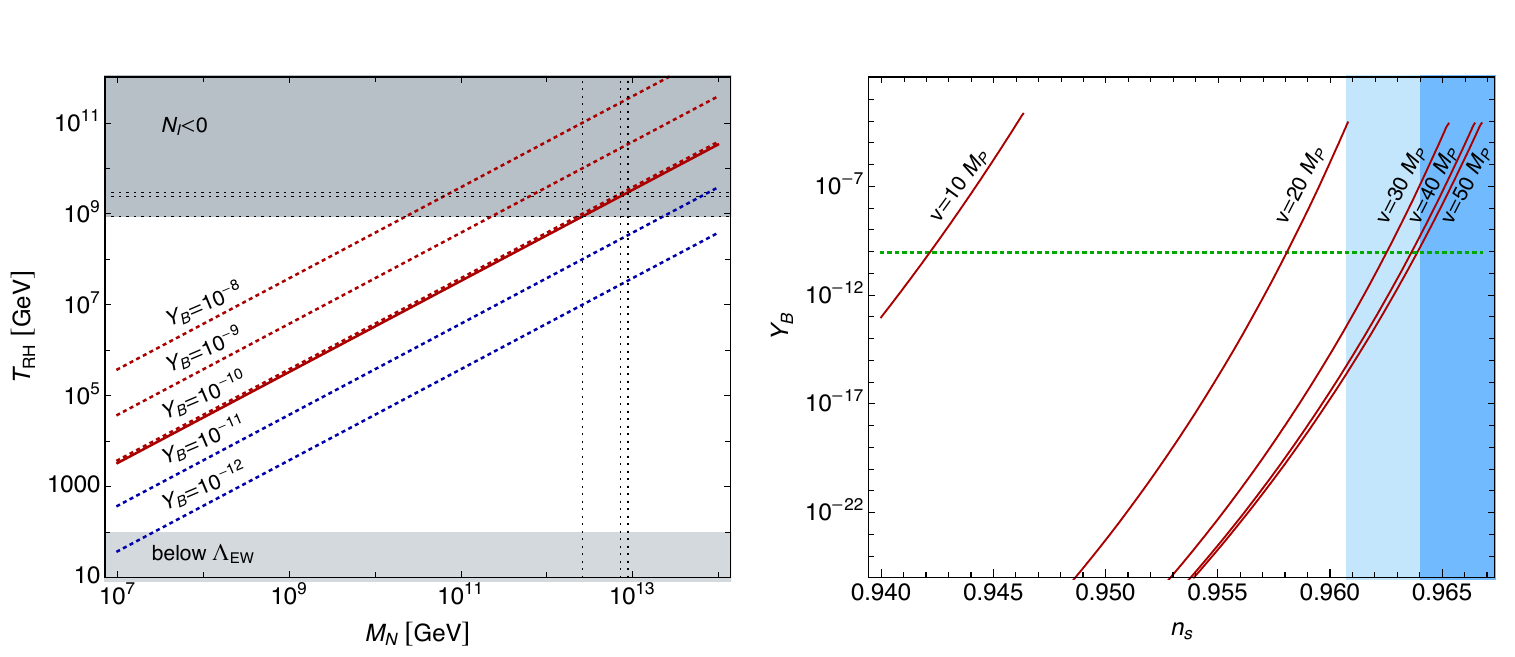}
\caption{The full parameter space in Coleman-Weinberg inflation with neutrino reheating in the $\Gamma_N^{}\ll \Gamma_\phi^{}$ scenario. In the left panel, horizontal and vertical dashed lines indicate the upper limits of $T_\mathrm{RH}^{}$ and $M_N^{}$ respectively, determined by different values of $v$. Other shadings are the same as in the figure~\ref{fig:poly2}.}
\label{fig:cw2}
\end{figure}

The full parameter space is shown in figures~\ref{fig:cw2} and \ref{fig:cw3}.
In the left panel of figure~\ref{fig:cw2}, the gray shaded region at the top is excluded by the requirement of $N_\mathrm{I}^{}>0$, which imposes an upper bound on $T_\mathrm{RH}^{}$ that varies with $v$. The dotted lines mark five $v$ values from $10 M_\mathrm{P}^{}$ to $50 M_\mathrm{P}^{}$, from bottom to top is the increasing order $v$. The upper three lines are nearly indistinguishable due to their close spacing. The same happens for the five vertical dotted lines, which show the kinetic upper limit of $M_N^{}$. The viable parameter space that produces sufficient baryon asymmetry $Y_\mathrm{B}^{}$ lies above the red line and below the dotted $T_\mathrm{RH}^{}$ limit. The $M_N^{}$ upper bounds are less restrictive than the $Y_\mathrm{B}^{}$ constraint. 

The right panel of figure~\ref{fig:cw2} displays the dependence of the baryon asymmetry $Y_\mathrm{B}^{}$ on the spectral index $n_s^{}$ with different values of $v$. We observe that $Y_\mathrm{B}^{}$ increases sharply with $n_s^{}$. The lower bounds of $n_s^{}$, determined by the requirement $N_k^{}\geq 29$, lie outside the plotted range. Using \eqref{eq:p-condition}, we derive and plot upper limits for $n_s^{}$. $v<20 M_\mathrm{P}^{}$ can hardly overlap with the $n_s^{}$ $1\sigma$ region. A larger $v$ corresponds to a larger $n_s^{}$. 

Figure~\ref{fig:cw3} presents our predictions for the tensor-to-scalar ratio. The dotted curves show the predicted $r$, with their solid segments indicating the $n_s^{}$ range capable of producing sufficient baryon asymmetry. The plot demonstrates that including the baryon asymmetry constraint  $Y_\mathrm{B}^{}$ substantially reduces the allowed parameter space. For a given spectral index $n_s^{}$, a larger $v$ corresponds to a larger $r$. Consequently, the allowed range of $v$ is constrained by $r$ from the top and $n_s^{}$ from the bottom. While showing no overlap with ACT-inclusive datasets, our results exhibit partial agreement with Planck-BK18 data for the parameter range $v\sim [13,47] M_\mathrm{P}^{}$.

\subsection{Example 4: Natural inflation}\label{sec:ni}
We consider natural inflation of the following form~\cite{Freese:1990rb,Adams:1992bn}
\begin{align}
V=\Lambda^4 \left(1+\cos\frac{\phi}{f}\right)\,.
\end{align}
The slow-roll parameters are
\begin{align}
\epsilon &=\displaystyle \frac{M_\mathrm{P}^2}{2f^2} \frac{1-\cos \chi}{1+\cos \chi}\,,\label{eq:ni-epsilon}\\
\eta &=\displaystyle -\frac{M_\mathrm{P}^2}{f^2} \frac{\cos \chi}{1+\cos \chi}\,,\label{eq:ni-eta}
\end{align}
where we introduce $\chi\equiv \phi/f$.
The e-folds number from the end of inflation to horizon exit is
\begin{align}
N_k^{} = \displaystyle - \frac{f^2}{M_\mathrm{P}^2}  \int_{\chi_e}^{\chi_k} \frac{1+\cos \chi}{\sin \chi} \mathrm{d}\chi 
= \frac{f^2}{M_\mathrm{P}^2} \mathrm{ln} \frac{1-\cos \chi_e^{}}{1-\cos \chi_k^{}}\,.\label{eq:ni-Nk}
\end{align}
$\epsilon(\chi_e^{})=1$ leads to 
\begin{align}
\cos \chi_e^{}=\frac{1-a}{1+a},\quad a\equiv \frac{2f^2}{M_\mathrm{P}^2}\,.\label{eq:ni-chie}
\end{align}
Substituting eq.~\eqref{eq:ni-chie} to eq.~\eqref{eq:ni-Nk}, we find
\begin{align}
\cos \chi_k^{} =1-b,\quad b\equiv \frac{2a}{1+a} e^{-N_k^{} \frac{M_\mathrm{P}^2}{f^2}} \,.\label{eq:ni-chik}
\end{align}
Putting $\chi_k^{}$ of eq.~\eqref{eq:ni-chik} in eqs.~\eqref{eq:ni-epsilon}, \eqref{eq:ni-eta}, we find
\begin{align}
\epsilon =\frac{M_\mathrm{P}^2}{2f^2} \frac{b}{2-b},\quad \eta=-\frac{M_\mathrm{P}^2}{2f^2} \frac{1-b}{2-b}\,,\label{eq:ni-epsilon2}
\end{align}
which leads to
\begin{align}
n_s^{}=1-6\epsilon+2\eta=1+\frac{M_\mathrm{P}^2}{f^2}\frac{b+2}{b-2}\,.\label{eq:ni-ns}
\end{align}
Since $b$ contains $N_k^{}$, we can express $N_k^{}$ in terms of $n_s^{}$ as 
\begin{align}
N_k^{}=-\left(\frac{f}{M_\mathrm{P}^{}}\right)^2 \mathrm{ln} \left[\left(1+\frac{M_\mathrm{P}^2}{2f^2}\right)\frac{1-n_s-\frac{M_\mathrm{P}^2}{f^2}}{1-n_s+\frac{M_\mathrm{P}^2}{f^2}}\right]\,.\label{eq:ni-Nk2}
\end{align}
The requirement of $N_k^{}$ being positive leads to a lower limit on $f$
\begin{align}
\frac{f}{M_\mathrm{P}^{}}> \frac{1}{\sqrt{1-n_s^{}}}\,.\label{eq:flimit}
\end{align}
Planck 2018 central value of $n_s^{}=0.9649$ gives $f>5.34 M_\mathrm{P}^{}$. A smaller $n_s$ renders a smaller $f$ lower limit. We can express $H_k^{}$ as a function of $n_s^{}$ with the help of eq.~\eqref{eq:ni-epsilon2} and eq.~\eqref{eq:ni-Nk2}. Similarly, we can get $V_\mathrm{end}^{}$ as a function of $n_s^{}$ with eq.~\eqref{eq:ni-chie} eq.~\eqref{eq:ni-chik} and $H_k^{}$. Note that all these quantities are also functions of $f$. The overall scale $\Lambda$ can be fixed by the scalar perturbation amplitude, which is at $\mathcal{O}(10^{16})$ GeV. 

\vspace{0.2cm}
\noindent{\bf Neutrino reheating predictions in the $\Gamma_N^{}\gg \Gamma_\phi^{}$ scenario}
\vspace{0.15cm}

\begin{figure}[t!]
\centering
\includegraphics[width=0.6\textwidth]{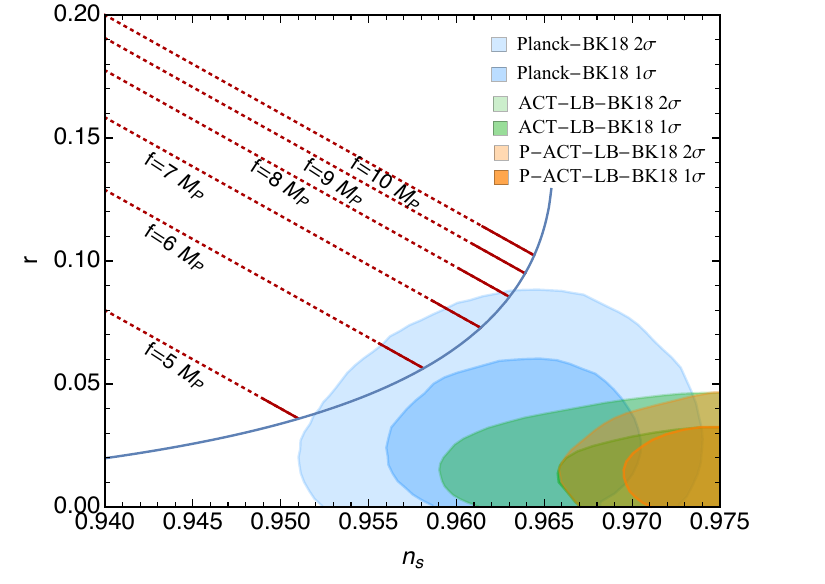}
\caption{Predictions in natural inflation with neutrino reheating. The red dotted lines show the predicted tensor-to-scalar ratio in the $\Gamma_N^{}\ll \Gamma_\phi^{}$ scenario, with their solid portions marking the range of spectral indices $n_s^{}$ that yield sufficiently large $Y_\mathrm{B}^{}$. The blue curve is the prediction in the $\Gamma_N^{}\gg \Gamma_\phi^{}$ scenario. Other shadings are the same as in figure~\ref{fig:poly3}.}
\label{fig:ni3}
\end{figure}

In this limit, $N_\mathrm{RH}^{}=0$ and the $N_\mathrm{RH}^{}$-$N_k^{}$ relation reduces to \eqref{eq:w=1/3}, from which we find a prediction of $f$ as a function of $n_s^{}$. It allows us to determine $M_\phi^{}$ and also the tensor-to-scalar ratio. $N_\mathrm{RH}^{}=0$ also allows us to equate $\rho_\mathrm{end}^{}$ with $\rho_\mathrm{RH}^{}$. Hence, we find a relation of $T_\mathrm{RH}^{}$ with $n_s$. The predictions are plotted on the $n_s^{}$ - $r$ plane in figure~\ref{fig:ni3} as the blue curve. The intersection with the Planck-BK18 $2\sigma$ region allows us to locate the viable parameter space:
\begin{align}
&f \in [5.09,8.26] M_\mathrm{P}^{} \nonumber \\
&T_\mathrm{RH}^{}\in [2.52,2.74]\times 10^{15} \mathrm{GeV} \nonumber\\
&M_\phi^{} \in [1.23,1.44]\times 10^{13} \mathrm{GeV}\,.\nonumber
\end{align}
The predicted baryon asymmetry are from $1.00\times 10^{-4}$ to $1.09\times 10^{-4}$, several orders larger than observed.

\vspace{0.2cm}
\noindent{\bf Neutrino reheating predictions in the $\Gamma_N^{}\ll \Gamma_\phi^{}$ scenario}
\vspace{0.15cm}

In this scenario, the duration of neutrino reheating $N_\mathrm{RH}^{}$ calculated from \eqref{eq:N1} and \eqref{eq:N2} depends on the five parameters: $M_\phi^{}$,~$M_N^{}$,~$T_\mathrm{RH}^{}$,~$n_s^{}$,~$f$. The inflaton mass is $M_\phi^{}=\Lambda^2/f$ with $\Lambda$ fixed by the amplitude of the scalar perturbation.  So, the four independent parameters are 
$$M_N^{},~T_\mathrm{RH}^{},~f,~n_s^{}\,.$$

\begin{figure}[t!]
\centering
\includegraphics[width=\textwidth]{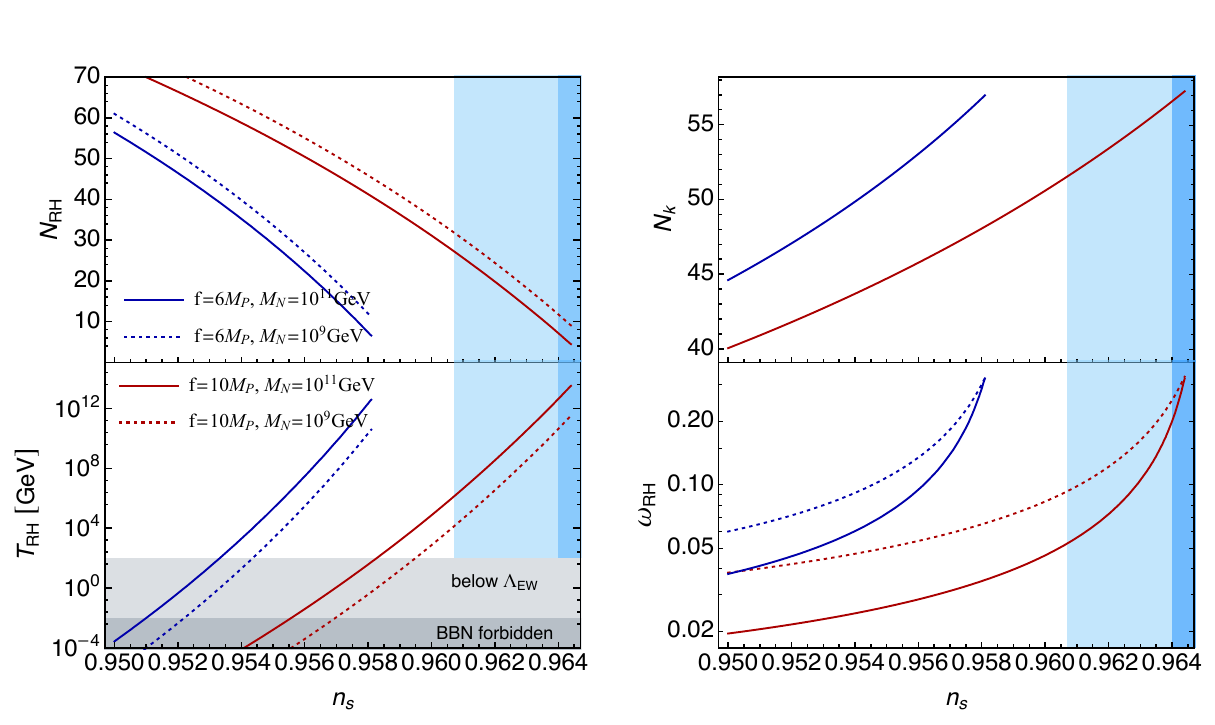}
\caption{Predictions in natural inflation with neutrino reheating in the $\Gamma_N^{}\ll \Gamma_\phi^{}$ scenario. Other shadings are the same as in the figure~\ref{fig:poly1}. }
\label{fig:ni1}
\end{figure}

As in previous examples, we first choose to work with $f=6 M_\mathrm{P}^{}, 10 M_\mathrm{P}^{}$ and $M_N=10^{11}, 10^9$ GeV. The results are shown in figure~\ref{fig:ni1}. We observe that the lines corresponding to $f=10 M_\mathrm{P}^{}$ overlap with the observed $n_s^{}$ $1\sigma$ region, whereas the lines corresponding to $f=6 M_\mathrm{P}^{}$ do not overlap at all. A larger $f$ corresponds to a lower $T_\mathrm{RH}^{}$, a smaller $N_k^{}$ and a smaller $\omega_\mathrm{RH}^{}$. Different $M_N^{}$ lines converge in the $\omega_\mathrm{RH}^{}$ - $n_s^{}$ plane. As the universe approaches radiation domination ($\omega=1/3$), the role of $M_N^{}$ becomes increasingly irrelevant. Our results confirm that requiring $\omega_\mathrm{RH}^{}\leq 1/3$ gives $r\geq 0.05$ \cite{Cook:2015vqa}, and are compatible with updated results in \cite{Martin:2013tda}.

\begin{figure}[t!]
\centering
\includegraphics[width=\textwidth]{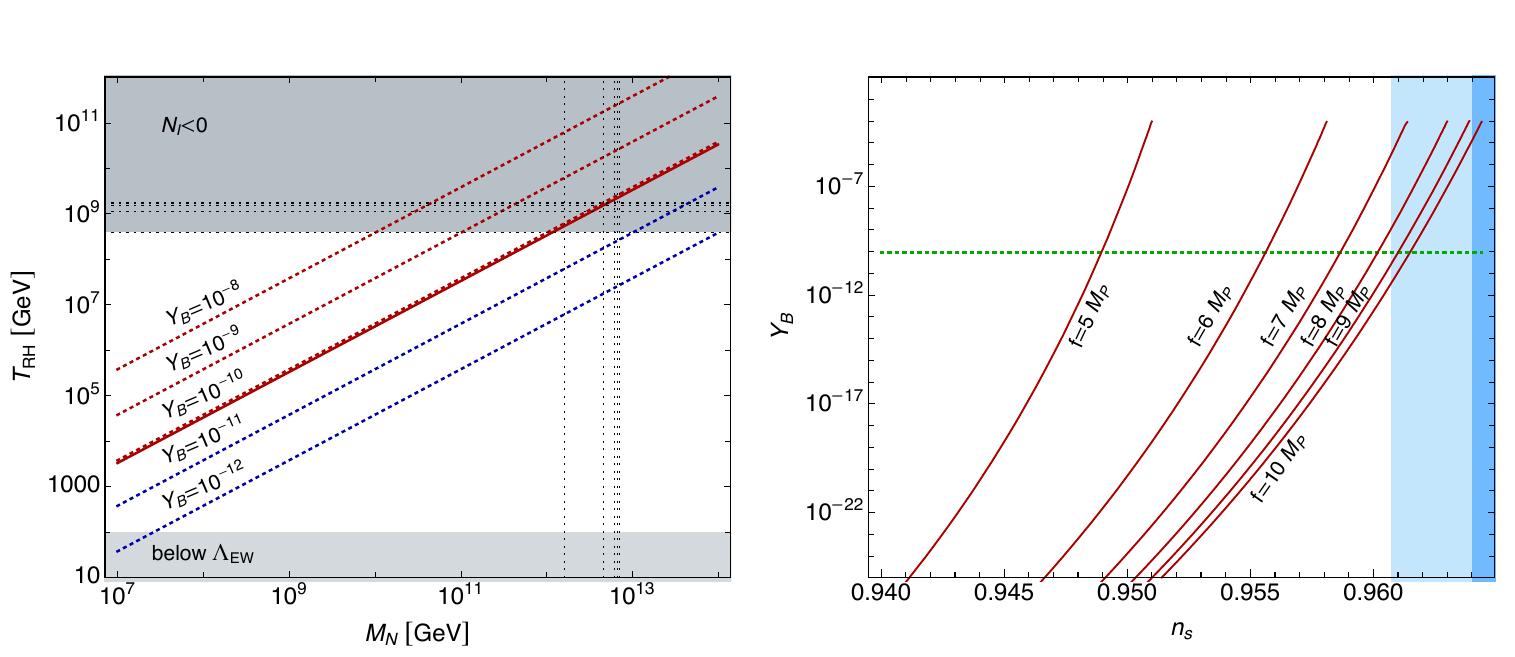}
\caption{The full parameter space in natural inflation with neutrino reheating in the $\Gamma_N^{}\ll \Gamma_\phi^{}$ scenario. In the left panel, horizontal and vertical dashed lines indicate the upper limits of $T_\mathrm{RH}^{}$ and $M_N^{}$ respectively, determined by different values of $f$. Other shadings are the same as in the figure~\ref{fig:poly2}.}
\label{fig:ni2}
\end{figure}

The complete parameter space is shown in figures~\ref{fig:ni2} and \ref{fig:ni3}. In the left panel of figure~\ref{fig:ni2}, model-dependent constraints appear as dashed lines representing upper bounds on both $T_\mathrm{RH}^{}$ and $M_N^{}$. The viable parameter space for generating sufficient baryon asymmetry corresponds to the upper-left white triangular region bounded by the red curve and the dashed $T_\mathrm{RH}^{}$ constraint.
In the right panel of figure~\ref{fig:ni2}, each red curve terminates at an upper bound determined by \eqref{eq:p-condition}. 
Combining the constraints from figures~\ref{fig:ni2} and~\ref{fig:ni3}, we identify the viable range for the parameter $f$ as $[5.09,8.26]~M_\mathrm{P}^{}$. This range predicts a sufficient baryon asymmetry and agrees with the Planck-BK18 $n_s^{}$-$r$ constraint at the $2\sigma$ confidence level. 
The degeneracy in $f$ can be broken by requiring the correct observed value of $Y_\mathrm{B}^{}$, as demonstrated in the right panel of figure~\ref{fig:ni2}.
While the predicted baryon asymmetry $Y_\mathrm{B}^{}$ favors larger values of $n_s^{}$, consistent with ACT data, including ACT constraint will exclude all natural inflation with neutrino reheating scenarios at $2\sigma$ confidence level.  In the natural inflation model, as $f$ grows, both $n_s^{}$ and $r$ grow. The tensor-to-scalar ratio is too large to be compatible with ACT data. 
Taking into account other effects such as dissipative effects~\cite{Reyimuaji:2020bkm,Montefalcone:2022jfw} or modified gravity ~\cite{Salvio:2019wcp,Salvio:2020axm,Simeon:2020lkd,Reyimuaji:2020goi,Zhang:2021ppy,Chen:2022dyq}
may suppress $r$ and render a natural inflation model compatible with constraints.

\section{Discussions and conclusions}\label{sec:conclusion}
In this work, we continue investigating the intriguing possibility of connecting inflation and neutrino physics through non-thermal leptogenesis~\cite{Zhang:2023oyo}. We assume a direct inflaton-RHN coupling. The RHNs decay reheats the universe through their subsequent decays to SM particles and we called it ``neutrino reheating". Depending on the relative sizes of $\Gamma_N^{}$ and $\Gamma_\phi^{}$, we consider two scenarios of neutrino reheating.
In the $\Gamma_N^{} \gg \Gamma_\phi^{}$ scenario, the RHNs decay instantaneously, corresponding to $N_\mathrm{RH}^{}=0$. We find predictions on $n_s^{}$, $T_\mathrm{RH}^{}$ and hence $Y_\mathrm{B}^{}$ for each model. The predicted $Y_\mathrm{B}^{}$ are at $\mathcal{O}(10^{-4})$ to $\mathcal{O}(10^{-5})$, which are much larger than the observed value and leave much space for model building.
In the $\Gamma_N^{} \ll \Gamma_\phi^{}$ scenario, there is an RHN matter dominant phase.
We calculate the duration of neutrino reheating.
Utilizing the $N_\mathrm{RH}^{}$-$N_k^{}$ relation, we establish a direct connection between the baryon asymmetry and the spectral index, bringing these two important observables onto the same plane (see right panels of figures~\ref{fig:poly2},~\ref{fig:star2},~\ref{fig:cw2},~\ref{fig:ni2}).

The direct link of $Y_\mathrm{B}^{}$ with $n_s^{}$ is only possible when we consider neutrino reheating constraints to inflation models. In ref.~\cite{Zhang:2023oyo}, we draw inflation-compatible lines in BAU parameter space. Inflationary constraints are considered under general assumptions with respect to reheating ($N_k^{}=60$). In this work, the inflationary observables are calculated via the $N_\mathrm{RH}^{}$-$N_k^{}$ relation, with neutrino reheating determined $N_\mathrm{RH}^{}$. In this context, $N_\mathrm{RH}^{}$ depends on $T_\mathrm{RH}^{}$, leading to a $T_\mathrm{RH}^{}$ - $n_s^{}$ relation, which in turn translates into a $Y_\mathrm{B}^{}$ - $n_s^{}$ relation, considering the analytical limits of non-thermal leptogenesis.

Reheating can break the degeneracy of inflation models. Neutrino reheating is particularly effective in this regard, as it includes predictions for $Y_\mathrm{B}^{}$, which are very sensitive to $n_s^{}$ and thus provide more restrictive constraints. In our considered inflation models, $Y_\mathrm{B}^{}$ changes rapidly with $n_s^{}$. Bringing $Y_\mathrm{B}^{}$ together with $(n_s^{},r)$ constrains the parameter space greatly as shown in figures~\ref{fig:poly3},~\ref{fig:star3},~\ref{fig:cw3},~\ref{fig:ni3}. 
Our results exhibit partial agreement with Planck-BK18 data for the parameter range $v\sim [13,47] M_\mathrm{P}^{}$ in the Coleman-Weinberg inflation, and $f\in [5.09,8.26]~M_\mathrm{P}^{}$ in the natural inflation model. 

The reheating temperature $T_\mathrm{RH}^{}$ is related to the spectral tilt through the $N_\mathrm{RH}^{}$-$N_k^{}$ relation with $N_\mathrm{RH}^{}$ determined by neutrino reheating. It is highly sensitive to $n_s^{}$ and spans several orders of magnitude given the $n_s^{}$ constraint. It also leads to rapid changes of the baryon asymmetry in the limit $\Gamma_N^{} \ll \Gamma_\phi^{}$. With the precise determination of $Y_\mathrm{B}^{}$, $T_\mathrm{RH}^{}$ is found to range from $10^4$ to $10^9$ GeV in the models considered. In the limit $\Gamma_N^{} \gg \Gamma_\phi^{}$, $T_\mathrm{RH}^{}$ is much higher, around $10^{15}$ GeV, which may lead to a potential gravitino problem in supersymmetric models~\cite{Khlopov:1984pf,Ellis:1984eq,Kohri:2005wn}. Possible solutions, other than lower $T_\mathrm{RH}^{}$, can be found in~\cite{Davidson:2008bu} and references therein.

Connecting inflation with neutrino physics through non-thermal leptogenesis with direct inflaton-RHN coupling offers a simple, natural, and testable framework describing the early univserse. The well-motivated components—inflation, type-I seesaw, and leptogenesis—provide straightforward and elegant solutions to several open questions in cosmology and particle physics. This approach naturally incorporates neutrino reheating, leaving no ambiguity regarding the early history of the universe. 
The predictions of neutrino reheating place $Y_\mathrm{B}^{}$ and $n_s^{}$ on the same plane, providing specific predictions that help break the degeneracy of inflationary models. There remains much to explore in this direction, including the incorporation of neutrino flavor structure and the refinement of the evolution process. We will address these topics in future work and look forward to upcoming cosmological and neutrino data to further test this framework.

\section*{Acknowledgements}
This work is supported by National Natural Science Foundation of China under grant No.12305116, Start-up Funds for Young Talents of Hebei University (No.521100223012).

\bibliographystyle{JHEP}
\bibliography{bibliography}

\end{document}